\documentclass[journal, 10pt]{IEEEtran}
\IEEEoverridecommandlockouts

\usepackage{amsmath}
\usepackage{mathtools}
\usepackage{cuted}
\usepackage{revsymb}
\usepackage[tight,footnotesize]{subfigure}
\usepackage{balance}
\usepackage{xcolor}
\usepackage{eqnarray}
\usepackage{commath}
\usepackage{graphicx}
\usepackage{textcomp}
\DeclareGraphicsExtensions{.eps}
\graphicspath{{./Figures/}}
\usepackage{multicol}
\usepackage{algorithm,algorithmic}
\usepackage{subfiles}
\usepackage{mathcomp}
\usepackage{subfig}
\usepackage{cite}
\usepackage{cases}
\usepackage{amssymb}

\usepackage{verbatim}
\normalsize

\usepackage[normalem]{ulem}

\begin{document}

\title{\ \\Selectivity of Protein Interactions Stimulated by Terahertz Signals}

 \author{Hadeel Elayan, Andrew W. Eckford, and Raviraj Adve 
%
%
%
\thanks{We would like to acknowledge the support of the National Science and Engineering Research Council, Canada, through its Discovery Grant program.}
\thanks{H. Elayan and R. Adve are with the Edward S. Rogers Department of Electrical and Computer Engineering, University of Toronto, Ontario, Canada, M5S 3G4 (e-mail: hadeel.mohammad@mail.utoronto.ca; rsadve@ece.utoronto.ca).}
\thanks{ A. Eckford is  with the Department of Electrical Engineering and Computer Science, York University, Ontario, Canada, M3J 1P3 (e-mail: aeckford@yorku.ca).}}

\maketitle
\begin{abstract} It has been established that Terahertz (THz) band signals can interact with biomolecules through resonant modes. Specifically, of interest here, protein activation. Our research goal is to show how directing the mechanical signaling inside protein molecules using THz signals can control changes in their structure and activate associated biochemical and biomechanical events.  To establish that,  we formulate a selectivity metric that quantifies the system performance and captures the capability of the nanoantenna to induce a conformational change in the desired protein molecule/population. The metric provides a score between~$-1$~and~$1$ that indicates the degree of control we have over the system to achieve targeted protein interactions. To develop the selectivity measure, we first use the Langevin stochastic equation driven by an external force  to model the protein behavior.   We  then determine the probability of protein folding by computing  the steady-state energy of the driven protein and  then generalize our model to account for protein populations.   Our numerical analysis results indicate that a maximum selectivity score is attained when only the targeted population experiences a folding behavior due to the impinging THz signal.  From the achieved selectivity values, we conclude that  the system  response not only depends on the resonant frequency but also on the system controlling parameters namely, the nanoantenna force, the damping constant, and the abundance of each protein population. The presented work  sheds light on the  potential associated with the electromagnetic-based control of protein networks, which could lead to a plethora of applications in the medical field ranging from bio-sensing to targeted therapy. 
\end{abstract}

%

 \section{Introduction}
 The engineering community is witnessing a new frontier
in the communication industry. A network infrastructure consisting of nanodevices is envisioned to enable robust, reliable, and coordinated data transmission. This will also allow  for myriad applications in several research fields, including bioengineering, wireless communication, nanotechnology, and environmental sciences.  Amongst others, in-vivo wireless nanosensor networks have been presented to provide fast and accurate disease diagnosis and treatment. These networks are capable  of operating inside the human body in real-time and will be of great benefit for medical monitoring and medical implant communication~\cite{shubair2015vivo}.

 Researchers have proposed various solutions to realize nanoscale communication, considering both molecular and electromagnetic (EM) communication paradigms. From an EM perspective, plasmonic nano-lasers, plasmonic nano-antennas  as well as single-photon detectors  all point to the Terahertz (THz) band, defined between 0.1 and 10 THz, as a key enabler of communication at the nanoscale. Numerical analysis and characterization of THz propagation through various body tissues have been presented in~\cite{7086348,abbasi2016terahertz}. A model that captures intra-body signal degradation is developed in~\cite{elayan2017terahertz}. A multi-layer system that accounts for the discrepancies  of human tissues for nano-biosensing applications can be found in~\cite{elayan2017multi}. Starting from these works, the importance of studying intra-body nanonetworks has emerged. 

THz vibrations are inherently involved in the functionality of biological systems since the energy scale of THz radiation is within the range of interactions between molecules~\cite{wei2018application}. This overlap in energy bands justifies the particular sensitivity of emerging THz techniques to the molecular motions that underlie intricate biological mechanisms. Specifically,  proteins are versatile macromolecules that are responsible for nearly all tasks in a cell. A critical function of proteins is their activity as enzymes, which are needed to catalyze almost all biological reactions. Regulation of the enzyme activity plays a key role in governing the cell behavior. One important feature associated with the protein structure is that it exhibits vibrational spectral features in the THz regime, corresponding to functionally relevant modes.  These modes are  regarded as the dynamics leading to  conformational  changes and biomolecular functions~\cite{markelz2008terahertz}. 

Protein mechanical vibrations can be dipole active, and thus probed using THz dielectric spectroscopy~\cite{acbas2014optical}. The extensive review in~\cite{markelz2008terahertz} explores measurements of the THz dielectric response on molecules. The author explains the contributions of the vibrational modes in providing possible information about protein conformational changes, ligand binding and oxidation state. Additionally, the authors in~\cite{carpinteri2017terahertz} and~\cite{carpinteri2019terahertz} investigated the mechanical vibrations of the protein lysozyme and the sodium-potassium protein membrane, respectively. Modal analysis was carried out by solving a multi-degree-of-freedom  vibration  problem and the results were experimentally verified using Raman spectroscopy. The  authors indicated a correlation between the resonant peaks found from experiments and the corresponding vibration modes given by  numerical simulations. 

Further, in a previous work, we proposed a stimuli-responsive paradigm which integrates EM and molecular communication by stimulating proteins in the human body. We analytically derived the mutual information and computed the capacity under different constraints for a two-state~\cite{elayan2020regulating} and a multi-state protein model~\cite{elayan2020information}. Nonetheless, a fundamental aspect that still must  be studied involves understanding the relation between the protein mechanical system and its statistical behavior. This entails being capable of determining the degree of control over the nanoantenna-protein interaction in inducing the desired functional conformational change.

In this work, we develop a theoretical framework to demonstrate the relationship between the protein's mechanical system and the probability of protein folding. We exploit the protein dynamics using  the Langevin stochastic equation exposed to an external force, where the protein molecule is modeled as a harmonic oscillator.  The  Langevin equation is used to capture the components of biological systems, which are constantly subject to  random Brownian motion~\cite{beyerle2019kinetics}. Since the aim is to model the impact of the nanoantenna on the protein states, we compute the steady-state energy of the induced protein motion and use it to obtain the corresponding Boltzmann's distribution. Our system is generalized to account for a protein population using a normal approximation to the binomial distribution.

Furthermore, we formulate a selectivity metric for both single and multiple proteins to serve as an indicator  that quantifies the system performance. The metric provides a score which captures  the  capability of the nanoantenna to stimulate a conformational change in the desired protein molecule/population without provoking  other untargeted proteins in the system to fold. The selectivity values range between $-1$ and $1$. A value of $-1$ indicates the worst-case scenario, where only the  undesired protein  molecule/population is  being provoked to fold. On the contrary, a value of $1$ indicates the best-case scenario, where only the desired protein molecule/population is being stimulated to fold. As such, the closer the  value achieved to is $1$, the higher the selectivity in the system. Finally, a value of $0$ reflects either the case where neither the desired nor the undesired proteins experience a conformational change or the case where both the desired and  undesired proteins experience  conformational changes. Both aforementioned scenarios indicate that the system is being non-selective since selectivity demands only the desired protein molecule/population to experience folding. From this perspective, we make the following  contributions:
\begin{itemize}
\item 
We formulate an expression for the induced protein steady-state energy  due to THz radiation. The energy expression links  the mechanical system  of the protein to the probability of protein folding.   
\item We generalize our system to account for a protein population. We then present a selectivity metric  to serve as a tool that determines the
capability of THz nanosensors in governing protein interactions
in an intra-body network. 
\item We  formulate a joint optimization problem to retrieve the optimal nanoantenna force and frequency values that maximize the selectivity. 
\end{itemize}

 The rest of the paper is organized as follows. In Sec.~\ref{Sec:Sec2}, we present the system model of the  driven protein dynamics. In Sec.~\ref{Sec:Sec3}, we formulate the  expression of  the steady-state energy driving the protein to change its conformation and use it to modify Boltzmann distribution. In Sec.~\ref{Sec:Sec4}, we formulate a metric to account  for the selectivity of the nanoantenna-protein interaction. In Sec.~\ref{Sec:Sec5}, we  demonstrate our numerical results based on the developed model. Finally, we draw our conclusions in Sec.~\ref{Sec:Sec6}. 
 \section{System Model}
 \label{Sec:Sec2}
 \subsection{Physical Basis}
  Our system is composed of a nanoantenna transmitter and a protein receiver. Plasmonic nanoantennas can provide both enhanced and controllable EM-matter interactions in addition to strong coupling between far-field radiation and localized sources at the nanoscale~\cite{aglieri2020improving}. These nanoantennas with typical dimensions on the order of a few hundred nanometers have found their applications in bio-sensing, bio-imaging, as well as energy harvesting. In this manuscript, our nanoantenna operates in the THz frequency range~\cite{rahm2015focus,manikandan2019structure}. 
 
THz plasmonic waves  have propagation distances of hundreds or thousands of wavelengths, whereas plasmonics at higher frequencies typically have propagation lengths of only a few tens of wavelengths~\cite{mittleman2013frontiers}. To read more about THz plasmonic waves, we  direct interested readers to this review article~\cite{zhang2020terahertz}.  It is to be noted that in our current work, we consider  a dipole  nanoantenna designed for the incident frequency being used, which is tuned to  the vibrational frequency of the desired protein, without considering the specifics of the antenna.

The transmission from the nanoantenna impinges EM waves on the  protein population  to target distinct vibrational modes and trigger a functional conformational change. Application of an external EM field to biological entities induces a relative redistribution of internal charges within the molecule with respect to the field lines~\cite{romanenko2017interaction}. This  results in a  dielectric response that is frequency dependent. Subsequently, the nanoantenna EM field leads to the polarization of the protein molecule. By examining the  protein dielectric response, dynamical transitions in  the protein structure at the THz frequency range can be attributed to  relaxational and resonant processes. Relaxational responses arise from the amino acid side chains, while resonant responses stem from the correlated motions of the protein structure~\cite{knab2006hydration}. Correlated motion occurs when many parts of the system are coupled and oscillate simultaneously.  This suggests the capability of tuning the nanoantenna to the vibrational frequency of the protein to provoke desired functional resonant interactions. 

The protein permittivity formulated as, $\varepsilon(\omega)=\varepsilon{'}(\omega)-j \varepsilon{''}(\omega)$, has its real part given by~\cite{knab2006hydration}
~\begin{equation}
\varepsilon{'}(\omega)=\varepsilon_{\infty}+\frac{ \varepsilon_{o}-\varepsilon_{\infty}}{1+(\omega \tau)^{2}}+\frac{(\varepsilon_{1}-\varepsilon_{\infty})\left[1-\left( \frac{\omega}{\omega_{o}} \right)^{2}\right]}{\left[ 1-\left( \frac{\omega}{\omega_{o}} \right)^{2} \right]^{2}+ (\omega\gamma)^{2}},
\end{equation}
while the imaginary part yields
\begin{equation}
\varepsilon{''}(\omega)=\frac{(\varepsilon_{o}-\varepsilon_\infty)\omega \tau}{1+(\omega \tau)^{2}}+\frac{(\varepsilon_{1}-\varepsilon_\infty)\omega \gamma}{\left[ 1-\left( \frac{\omega}{\omega_{o}} \right)^{2} \right]^{2}+ (\omega \gamma)^{2}}.
\end{equation}
Here, $\tau$ is the protein relaxation time, $\omega_o$ is  the natural   frequency of the protein, and $\gamma$ is the damping constant. $\varepsilon_{\infty}$ is the permittivity at the high frequency limit, $\varepsilon_{o}$ is the relative permittivity at low frequencies (static region) and $\varepsilon_{1}$ refers to an intermediate permittivity value.  The damping constant $\gamma$ governs the magnitude of the resonances. In fact,  damping forces, which slow the motion of proteins, are due to both solvent and protein friction, where the viscosity arises either from the surrounding fluid or from interactions between amino acids. 

 According to structural mechanics, if an external harmonic excitation has a frequency which matches one  of the natural frequencies of the system, then resonance occurs, and the vibrational amplitude of the structure increases~\cite{bassani2017terahertz}.  When two oscillators are in resonance, two important characteristics hold, namely, high energy efficiency and rapidity. This means that nearly all of the energy can be transferred between the elements and that the speed of resonant exchange in energy is fast~\cite{chou1977biological}. As such, a  nanoantenna tuned to the resonant frequency of a distinct protein vibrational mode can stimulate protein signaling. By triggering the protein to adopt a folding behavior,  a series of downstream events occur within the cell, enabling a functional response.

To study and evaluate the impact of the nanoantenna-protein interaction, we formulate a \textit{selectivity} metric  to show the capability of the presented system in discriminating the desired response from adjacent inputs. Particularly, the devel-
oped metric measures the capability of the nanoantenna to stimulate the desired protein molecule/population to change its conformation without impacting the conformation of other proteins in the system. Such a metric provides a powerful tool as it gives a single value to quantitatively decide whether or not an interaction is sufficiently controllable.

The selectivity analysis lays the grounds for directed signaling, where desired proteins are driven  towards a conformation that evokes a particular response. It also paves the path towards a plethora of applications in the medical field. On the one hand, it allows us to recognize  mechanisms for selectively targeting proteins  involved in carcinogenesis, a procedure which can be utilized for the early diagnosis of cancer cells. On the other hand, it could further the understanding of neurodegenerative diseases, which are primarily caused due to the aggregation of misfolded proteins.   Finally, the presented work has broad implications for improving our understanding of proteins, protein engineering and better drug design. 

\subsection{Mathematical Model}
 \label{Sec:math_mod}

To model the protein  dynamics,  we use the Langevin stochastic equation under the influence of an external force. The Langevin equation allows us to  analyze the forces acting on the protein by virtue of the impacts received from the nanoantenna external force as well as the random forces originating from the interactions of the surrounding particles. It is given as~\cite{nadler1987molecular} \begin{equation}m \frac{d^2 x}{dt^2}+\beta \frac{dx}{dt}+kx(t)=f_{ex}(t)+f_{\zeta}(t), \label{eq:main_0}
\end{equation}
where $x=x(t)$ is the protein coordinate, $m$ is the protein mass, $k$ is the protein spring constant (stiffness), and $\beta$ is the damping coefficient. In~\eqref{eq:main_0}, $f_{ex}(t)=f_o\cos(\omega t)$ is the external driving nanoantenna force. This force arises from the interaction of the incident electric field of the antenna with the protein, where the protein is treated as a dielectric. Mathematically, the force scales linearly with the gradient squared of the electric field~\cite{seyedi2018protein}. 

In addition, $f_{\zeta}(t)$ is the stochastic force acting on the protein, which is governed by a white-noise fluctuation-dissipation relation as follows~\cite{balakrishnan2008elements}
\begin{equation}
f_{\zeta}(t)={\sqrt{\Gamma}}\zeta(t).
\label{eq:s_f}
\end{equation}
$\zeta(t)$ is a white, Gaussian random process with moments~\cite{balakrishnan2008elements}
\begin{equation} 
\begin{split}
\langle \zeta(t_1)\rangle&=0 \\
\langle \zeta(t_1)\zeta(t_2)\rangle&=\delta(t_1-t_2).
\label{eq:moments}
\end{split}
\end{equation}
 Here, $\left\langle \cdot \right\rangle$ defines the statistical average of a quantity. In~\eqref{eq:s_f}, $\Gamma= 2\gamma mk_bT$ denotes the strength of the noise $\zeta(t)$. It fixes the amplitude of the fluctuation in the random force in terms of both the temperature  and the dissipation coefficient~\cite{balakrishnan2008elements}.  In addition,~\eqref{eq:main_0} can be re-written as
 \begin{equation} \frac{d^2 x}{dt^2}+ \gamma\frac{dx}{dt}+\omega_o^2x(t)=\frac{1}{m}\left[ f_{ex}(t)+f_{\zeta}(t) \right],\\ \label{eq:main_1}
\end{equation}where $\gamma={\beta}/{m}$ and $\omega_o^2=k/m$. 
  
Green's function is a powerful mathematical tool to solve inhomogeneous differential equations. It provides the solution to the inhomogeneous equation with a forcing term given by a point source. For an arbitrary forcing term, the  solution to the same equation is formulated  by integrating Green's function against the forcing term. The frequency-domain solution of Green's function corresponding to~\eqref{eq:main_1} is found as~\cite{byron2012mathematics} 
\begin{equation}
G(\omega,t')=\frac{e^{j\omega t'}}{m\left[ -\omega^2- j\gamma \omega+\omega^2_o \right]}.
\label{eq:gfd}
\end{equation}
 The final form of Green's function is given by~\cite{byron2012mathematics}
\begin{equation}G(t,t')=\frac{1}{m\alpha}e^{-\frac{\gamma}{2}(t-t')}\sin\left[ {\alpha} (t-t') \right]\Theta (t-t'),
\end{equation}
where $\alpha= \sqrt{\omega^2_o-\frac{\gamma^2}{4}}$ and $\Theta (t-t')$ denotes the step function
\begin{equation}
\Theta (t-t')=\bigg\{\begin{array}{c}
1,\quad\quad\text {for} \,\,\,t\geq t'\ \\
0,\quad\quad \text{otherwise.} \\
\end{array}
\end{equation}

In this paper, we will consider $\omega_o>\frac{\gamma}{2}$. Physically, this condition satisfies protein collective vibrations, where protein modes in  the THz frequency  range have been shown to be underdamped even in aqueous solutions~\cite{turton2014terahertz,liu2008studies,deng2021near}.  

The solution of~\eqref{eq:main_0} can be decomposed into two parts, one related to the deterministic nanoantenna external force $f_{ex}(t)$ and the other related to the stochastic force $f_{\zeta}(t)$, resulting in $x(t)=x_{ex}(t)+x_{\zeta}(t)$.
The solution  corresponding to the nanoantenna force can be expressed using the time-domain Green's function as~\cite{yaghoubi2017energetics}
\begin{equation}
\begin{split}
x_{ex}(t)&={\int_{0}}^{\infty}G(t,t')f_{ex}(t')dt' \\
&=\frac{f_o }{m\alpha}{\int_{0}^t}e^{-\frac{\gamma}{2}(t-t')}{\sin}\left[ \alpha(t-t') \right]\cos(\omega t')dt' \\
&=A\cos(\omega t- \phi)-A\frac{\omega_o}{\alpha}\cos(\omega t-\varphi)e^{-\frac{\gamma }{2}t},
\end{split}
\label{eq:disp_main}
\end{equation}
where 
\begin{equation}
A=\frac{f_{o}}{m\sqrt{(\omega_o^2-\omega^2)^2+(\omega \gamma)^2}},
\label{eq:amplitude}
\end{equation}

\begin{equation}
\tan \phi=\frac{\gamma \omega}{{(\omega^2_o-\omega^2)}},
\label{eq:phase_tan}
\end{equation}
and
\begin{equation}
\tan \varphi=\frac{\gamma(\omega_o^2+\omega^2)}{{2\alpha(\omega^2_o-\omega^2)}}.
\end{equation}
Similarly, the solution  corresponding to the stochastic component can be expressed as
\begin{equation}
\begin{split}
x_\zeta(t)&=\int_{0}^{\infty}G(t,t')f_{\zeta}(t')dt' \\
&=\sqrt{\Gamma}\int^{\infty}_{0}G(t,t')\zeta(t')dt'. 
\end{split}
\label{eq:pos_n}
\end{equation}
Here $x_\zeta(t)$ is a linear function of $\zeta(t)$, and therefore  is also Gaussian.

\section{Boltzmann Distribution: Relating Energy to Probability}
\label{Sec:Sec3}
Amino acids are considered the building blocks of protein synthesis.  The amino acids that reside in the cytoplasm conduct movements through the rotation of the intracellular churn alongside the molecular interactions upon them. In addition,  ATP usage serves as a biological instigator for the interactions of amino acids. Such active and mass consumption  of  amino acids necessitate the constant flow of high levels of energy, which in absence results in further irreparable harm to cellular functions by disabling the conduction of protein activities~\cite{orhan2016stimulation}.

The necessity of energy exchange to trigger a protein conformational change in our system highlights the need for interfacing THz-band signals with protein molecules to initiate resonance. This will allow  minuscule bodies  to express large amplitude oscillations. To find the  total energy of a protein stimulated by an external force in a stochastic environment, we compute both the kinetic and potential energy  contributions as follows~\cite{elayan2021enabling}
\begin{equation}
\left\langle E_{tot} \right\rangle_{ss}=\frac{1}{2}m \left\langle v^2(t) \right\rangle_{ss} +\frac{1}{2}k \left\langle x^2(t) \right\rangle_{ss}.
\label{eq:driving_energy}
\end{equation}We denote by $\left\langle\cdot\right\rangle_{ss}$ the  statistical  average value in steady-state, where all transient effects die out. We also use $\overline{\left(\cdot\right)}_{ss}$ to refer to the mean value in  steady-state. The only difference between both notations  is that one of them comes from  a stochastic component, while the other comes from a deterministic component, respectively. To derive~\eqref{eq:driving_energy}, we utilize  the solution of the  Langevin equation derived in Sec.~\ref{Sec:math_mod}, which is given by $x_{ex}(t)$ and $x_{\zeta}(t)$. 

We first evaluate the potential energy by finding the mean squared   displacement of the protein, $\left\langle x^2(t) \right\rangle_{ss}$, which is decomposed into the contributions of the external force and stochastic force, respectively. It is given as 
\begin{equation}
\left\langle x^2(t) \right\rangle_{ss}=\overline{\left(x_{ex}^2(t)\right)}_{ss}+\left\langle x_{\zeta}^2(t)\right\rangle_{ss}.
\label{eq:x_ss}
\end{equation}
$\left\langle x^2(t) \right\rangle_{ss}$  is a measure of the spatial extent of the protein motion. From~\eqref{eq:disp_main}, we compute $\overline{\left(x_{ex}^2(t)\right)}_{ss}$, as\begin{equation}
\begin{split}
\overline{\left(x_{ex}^2(t)\right)}_{ss}&=\overline{\left( \left(A\cos(\omega t- \phi)-A\frac{\omega_o}{\alpha}\cos(\omega t-\varphi)e^{-\frac{\gamma }{2}t} \right)^2 \right)}_{ss}\\&=\frac{A^2}{2}.
\end{split}
\end{equation}  In addition, to find  $\left\langle x_{\zeta}^2(t)\right\rangle_{ss}$, we apply Parseval's theorem, where  we use the following corollaries
\begin{equation}
\left\langle x^2_{\zeta}(t) \right\rangle_{ss}=\frac{1}{T}\int_{-T/2}^{T/2}\left\langle|x_{\zeta}(t)|^2\right\rangle dt=\frac{1}{2\pi}\int_{-\infty}^{\infty}\left\langle|x_{\zeta}(\omega)|^2 \right\rangle  d\omega.
\label{eq:long}
\end{equation}$\left\langle|x_{\zeta}(\omega)|^2\right\rangle$ is found from~\eqref{eq:gfd} using the frequency domain Green's function along with the stochastic force defined in~\eqref{eq:s_f} as follows
 \begin{equation}
\begin{split}
\left\langle|x_{\zeta}(\omega)|^2\right\rangle=|G(\omega)\sqrt{\Gamma}|^2= \frac{\Gamma}{m^2\bigg[ \left(\omega_{o}^2-\omega^2)^{2}+(\omega\gamma\right)^2 \bigg]} 
\end{split}.
\label{eq:form_1}
\end{equation} 
From~\eqref{eq:form_1}, we compute the integration in~\eqref{eq:long} by using the residue theorem, which  evaluates line integrals of analytic functions over closed curves. This results in
\begin{equation}
\begin{split}
\left\langle x^2_{\zeta}(t) \right\rangle_{ss}&=\frac{1}{2\pi}\int_{-\infty}^{\infty}\left\langle|x_{\zeta}(\omega)|^2 \right\rangle d\omega  \\
&=\frac{1}{2\pi}\int_{-\infty}^{\infty}\frac{\Gamma}{m^2\bigg[(\omega_{o}^2-\omega^2)^{2}+(\omega\gamma)^2\bigg]}
d\omega\\
&=\frac{\Gamma}{2\pi m^{2}}\frac{ \pi}{\gamma\omega_{o}^2} =\frac{k_{b}T}{m\omega^2_o}=\frac{k_{b}T}{k}.
\end{split}
\label{eq:disp_0}
\end{equation}

In a similar manner, to evaluate the kinetic energy, we need to compute $\left\langle v^2(t) \right\rangle_{ss}$  which is given as
\begin{equation}
\left\langle v^2(t) \right\rangle_{ss}=\overline{\left(v_{ex}^2(t)\right)}_{ss}+\left\langle v_{\zeta}^2(t)\right\rangle_{ss}.
\label{eq:v_ss}
\end{equation}
$\overline{\left(v_{ex}^2(t)\right)}_{ss}$ is calculated from~the derivative of the displacement $x_{ex}(t)$ given in~\eqref{eq:disp_main}, where we first find $v_{ex}(t)$ as  follows
\begin{equation}
\begin{split}
 v_{ex}(t) &=  \dot x_{ex}(t) =-\omega A\sin(\omega t -\phi)\\&+A\frac{\omega_o}{\alpha}e^{-\frac{\gamma }{2}t}\left(\omega\sin(\omega t-\varphi)+\frac{\gamma}{2}\cos(\omega t-\varphi) \right).
\label{eq:velo_ex}
\end{split} 
\end{equation} 
We then compute the mean squared value of~\eqref{eq:velo_ex}  which leads to\begin{equation}
\overline{\left(v_{ex}^2(t)\right)}_{ss}= \overline{\left( \left( \dot x_{ex}(t) \right)^2 \right) _{ss}}=\frac{\omega ^{2}A^{2}}{2}.
\label{eq:part1}
\end{equation}
Likewise, $\left\langle v_{\zeta}^2(t)\right\rangle_{ss}$ is found from the Fourier transform of the derivative of the displacement. From~\eqref{eq:disp_0}, we have 
\begin{equation}
\begin{split}
\left\langle v^2_{\zeta}(t)\right\rangle_{ss}&=\frac{1}{2\pi}\int_{-\infty}^{\infty}\left\langle|x_{\zeta}(\omega)|^2\right\rangle\omega^2 d\omega\\
&=\frac{\Gamma}{2\pi m^{2}}\int_{-\infty}^{\infty}\frac{\omega^2}{\left(\omega_{o}^2-\omega^2)^{2}+(\omega\gamma\right)^2 }
d\omega \\
&=\frac{\gamma k_bT}{\pi m}\frac{\pi}{\gamma} \\
&=\frac{k_bT}{m}.
\end{split}
\label{eq:part2}
\end{equation}
Finally, we can find the total steady-state energy of the driven protein motion by substituting~\eqref{eq:x_ss} and~\eqref{eq:v_ss} in~\eqref{eq:driving_energy} yielding
\begin{equation}
\begin{split}
\left\langle E_{\mathrm{tot}}\right\rangle _{ss}&=\frac{1}{4}mA^2(\omega^2+\omega^2_o)+k_{b}T\\
&=\underbrace{\frac{1}{4}\bigg[ \frac{f^2_{o}(\omega^2+\omega^2_o)}{m((\omega_o^2-\omega^2)^2+(\omega \gamma)^2)}\bigg]}_{\text{THz Force Contribution}} +\underbrace{k_{b}T}_{\text{Noise Effect}}.
\end{split}
\label{eq:average_energy}
\end{equation}From~\eqref{eq:average_energy}, we can see that the derived expression reflects the capability of the applied electric field since the  energy absorbed is in excess of that of  thermal noise, i.e., $k_bT$. Thus, for resonances to have an important effect on biological systems, the effect of the applied field should be greater than the corresponding random field.  

Moreover, signaling inside proteins results in a spring-like effect which shifts their energy~\cite{orr2006mechanisms}. Protein structures are therefore investigated using energy functions where they obey statistical laws based on the Boltzmann distribution. Specifically, the Boltzmann distribution  provides the probability that a system will be in a certain state as a function of the state's energy and system temperature~\cite{finkelstein1995protein}. In the presented two-state model, the protein resembles a binary biological switch. The states of the protein depicted are folded and unfolded, as  those govern the activation of biological  processes and chemical interactions. The input to our  channel is the force induced by the nanoantenna, while  the output is the state of the protein. 

The rate of protein folding is given by~\cite{sachs1991stochastic}
\begin{equation}
r_{f}=r_0\exp \left(\frac{-E_f}{k_bT}\right),
\label{eq:fr0}
\end{equation}
where $E_f$ denotes the Gibbs free energy generated through metabolic processes and associated with the folded protein state. $r_0$ is a scale factor that preserves detailed balance. In our case, the rate of protein folding must incorporate the steady-state energy of the driven  protein motion, $\left\langle E_{\mathrm{tot}}\right\rangle _{ss}$, computed in~\eqref{eq:average_energy}  as follows\begin{equation}
r_{f}=r_0\exp \left(\frac{-E_f+\langle E_{\mathrm{tot}}\rangle_{ss}}{k_bT}\right).
\label{eq:fr}
\end{equation}
In addition, the unfolding transition is a relaxation process that returns
the protein to the unfolded state. In this view, the unfolded state is actually an equilibrium
distribution of many unfolded or partially folded conformational states. Such a
process is considered independent of the imposed signal~\cite{anfinsen1973principles} and is given by
\begin{equation}
r_{u}=r_0\exp \left(\frac{-E_u}{k_bT}\right),
\label{eq:br}
\end{equation} where $E_u$ denotes the Gibbs free energy associated with the unfolded protein state. Consequently, the rate of change of the protein folding probability is 
\begin{equation}
\frac{d}{dt}p_{f}(t)=-r_{u}p_{f}(t) +r_{f} (1-p_{f}(t)),
\label{eq:master_eq}
\end{equation}
where $p_{f}(t)$ is the probability that
the protein is in the folded  state at time $t$. The corresponding probability of the unfolded state is given by
$p_{u}(t)=1-p_{f}(t)$. From (\ref{eq:master_eq}), the stationary solution can be found as
\begin{equation}
p_{F}=\frac{r_f}{r_f+r_u}.
\label{eq:pf}
\end{equation}
Substituting \eqref{eq:fr} and \eqref{eq:br} in \eqref{eq:pf} yields 
\begin{equation}
p_{F}=\frac{1}{1+\exp\left(\frac{\Delta E-\langle E_{\mathrm{tot}}\rangle_{ss}}{k_{b}T}\right)},
\label{eq:pf_ff}
\end{equation}
where $\Delta E=E_f-E_u$. The values of $\Delta E$ for different proteins can be found from experimental studies available in the literature since protein folding and unfolding is a naturally occurring phenomenon driven by the change in free energy.
 
 The state of a single protein  can be regarded as
a Bernoulli random variable with a probability of success, $p_F$. As such, when considering a system composed of $n$ proteins, the random number of folded proteins, $n_F$, follows a binomial distribution that is given by\begin{equation}
p(n_F|p_{F})=\left( \begin{array}{c}
n\\
n_F\\
\end{array} \right)p_{F}^{n_F}(1-p_{F})^{n-n_F}.
\end{equation}
In addition, due to the large diversity of proteins seen in an intra-body environment, the binomial distribution can be approximated by a normal distribution with mean, $\mu=np_{F}$, and variance, $\sigma^2=np_{F}(1-p_{F})$, i.e.,
\begin{equation}
n_{F}\sim\mathcal{N}\left(np_{F}, np_{F}(1-p_{F})\right).
\end{equation}
We define protein population as the number of copies of a protein molecule in a cell. Hence, the number of folded proteins of each  protein population, $i$, can be expressed as a normal distribution as follows 
\begin{equation}
n_{F,i}\sim\mathcal{N}\left(np_{F,i}, np_{F,i}(1-p_{F,i})\right).
\end{equation}

\section{ Selectivity of the Nanoantenna-Protein Interaction}
\label{Sec:Sec4}
  Folding of proteins into their correct native structure is key to their function, whereas unfolded or misfolded proteins contribute to a pathology of many diseases~\cite{sweeney2017protein}. When a  nanoantenna targets a protein population, it is expected to affect the conformational change in the cytoplasmic domain of the \emph{desired} receptor molecules in order to induce a particular functional response.  In other words, a protein  interaction must be selective to achieve control over the protein network. 
  
  In the context of this work, \textit{selectivity} is introduced as a system performance metric that quantifies  the ability of the nanoantenna to provoke a conformational change  in the desired protein molecule/population  without impacting  a similar conformation of other untargeted proteins. A maximum selectivity score is expected when only the targeted population experiences a folding behavior due to the impinging THz signal. Therefore, the selectivity is an indicator as to whether  or not an interaction is sufficiently controllable. The focus on developing a selectivity metric stems from the fact  that  selectivity is the most important feature that reflects if  THz-stimulated interactions can be adequately controlled to  enable useful signal transduction processes. 

Selectivity arises in the literature in various scenarios. In drug discovery, it refers to the experimenter's ability to design  maximally selective ligands that act on specific targets. Poor selectivity in this context may result in toxicity and side effects~\cite{bosc2017use}. Moreover, for ion channels, the selectivity feature arises at permeable membranes, where it allows the passage of some molecules and inhibits the passage  of others~\cite{bostick2007selectivity}. Further, selectivity appears as a measure of the performance of a radio receiver when it responds only to the radio signal it is tuned to and rejects other signals nearby in frequency.

   In this work, the presented selectivity metric has been formulated after extensively investigating the literature. Our  choice of  measure relies on the interpretation of the metric in terms of the problem considered, its analytical properties
and ease of computation. In fact, it came to our attention that many of the available measures used in other contexts and applications, such  as the Bhattacharyya distance,  the Chernoff distance, and the Kullback-Leibler divergence, are unbounded from above, and therefore are not suitable for our application since they cannot directly guide experimenters to whether an interaction is selective or not.

 To assess the level of  control imposed by THz signals over protein interactions,  we  propose in this work a selectivity metric given by
\begin{equation}
S({p_{F,d},p_{F,ud}})=\frac{p_{F,d}-p_{F,ud}}{\max (p_{F,d},p_{F,ud})},
\label{eq:metric1}
\end{equation}
where $p_{F,d}$ is the probability of the folded state of the desired protein, while $p_{F,ud}$ is the probability of the folded state of the undesired protein. The denominator involves a  $\max$ function for normalization purposes. Since the developed metric relies on the stationary probability difference, $p_{F,d}-p_{F,ud}$, it  captures both the  system mechanical parameters and the amount of energy stored in the THz-stimulated protein.

Our developed selectivity metric has a  number of properties which renders it  as a powerful tool for evaluating protein interactions in the system. Those properties include:
\begin{itemize}
\item 
$S({p_{F,d},p_{F,ud}})$ is bounded, such that $-1\leq\ S(p_{F,d},p_{F,ud}) \leq 1$. 
\item $S(p_{F,d},p_{F,ud})$ has a \textbf{\textit{maximum value}} of 1 indicating that \textbf{both} the targeted protein was  selected \textbf{and }the untargeted protein  was  not selected. This occurs when $p_{F,d}=1$ and $p_{F,ud}=0$.  A value of 1 indicates the best-case scenario.  
\item $S(p_{F,d},p_{F,ud})$ has a \textbf{\textit{minimum value}} of -1 indicating that \textbf{both} the untargeted protein was  selected \textbf{and }the targeted protein was  not selected. This occurs when $p_{F,d}=0$ and $p_{F,ud}=1$. A value of -1 indicates the worst-case scenario. 

\item $S(p_{F,d},p_{F,ud})$ is continuous and an increasing function of $p_{F,d}$ when $p_{F,ud}$ is fixed. 
 \end{itemize}
 
When dealing with a  protein \textit{population}, the selectivity metric can be adjusted based on the separation of the means of the populations. The comparison of two independent population means is a very common tool. It actually provides a way to test whether the two groups differ  by capturing the overlap between the protein populations in the system. As such, it can signal whether  or not it is possible  to target the desired protein population. In this case, the selectivity metric is defined as
\begin{equation}
S({\mu_d}, \mu_{ud})=\frac{\mu_d-\mu_{ud}}{\max (\mu_d,\mu_{ud})},
\label{eq:metric2}
\end{equation}
where $\mu_d$ and $\mu_{ud}$ are the means of the desired and undesired protein populations, respectively. The same properties which hold for~(\ref{eq:metric1}) still holds in~(\ref{eq:metric2}). In fact, when the population of both the desired and undesired protein categories equal $n$,~(\ref{eq:metric2}) reduces to~(\ref{eq:metric1}), where  $S(\mu_d,\mu_{ud})= S(np_{F,d},np_{F,ud})=S(p_{F,d},p_{F,ud})$.

 In addition, an interesting feature of the presented measure is its scalability as it can  account for multiple undesired proteins in the system, while still holding its  aforementioned properties.  For such a case, the selectivity is given by
 \begin{equation}
S({\mu_d,\boldsymbol{\mu_{ud}}})=\frac{\mu_d-\sum_i\mu_{ud,i}}{\max (\mu_d,\sum_{i}\mu_{ud,i})},
\end{equation}
where $\sum_i\mu_{ud,i}$ refers to the sum over all the means of the undesired protein populations in the system.
\subsection{Optimal System Design Parameters}
\label{Sec:Sec4a}
In order to trigger a specific protein population to activate the desired system response,  we must ensure that the system achieves maximum selectivity. From~\eqref{eq:average_energy}, we can notice that  the nanoantenna frequency, $\omega$, and the nanoantenna force, $f_o$, are the only system parameters that we can control. Thereby, we formulate a joint optimization problem to maximize the selectivity  with respect to those parameters. To simplify~\eqref{eq:metric1}, an intuitive assumption is to only consider the scenario $p_{F,d}~>~p_{F,ud}$. This holds true since we  are tuning the nanoantenna to the resonant frequency of the desired protein population. Thus, the selectivity  becomes 
\begin{equation}
\begin{split}
S(p_{F,d},p_{F,ud})&=\frac{p_{F,d}-p_{F,ud}}{p_{F,d}} \\
&= 1-\frac{p_{F,ud}}{p_{F,d}}.  \\
\end{split}
\end{equation}
Hence, what we are essentially interested in maximizing is the stationary folding probability ratio of the desired and undesired proteins, ${p_{F,d}}/{p_{F,ud}}$. The optimization problem is therefore given as 
\begin{equation}
\begin{aligned}
& \underset{f_o,\text{$\omega$}}{\text{maximize} \,\,}
& & \text{$S(p_{F,d},p_{F,ud})$} \\
& \text{subject to}
& & 0\leq p_{F,d}\leq\ 1 \\
& & & 0\leq p_{F,ud}\leq\ 1. \\
\end{aligned}
 \label{eq:sys_const1}
\end{equation}
 To solve \eqref{eq:sys_const1}, we use  the fixed-point iteration method~\cite{cegielski2012iterative}. Specifically, given a function $g(x)$ defined on  real numbers,  and given a point $x_0$  in the domain of $g(x)$, the fixed-point iteration is $x_{j+1}=g(x_j), j=0,1,2...$, which gives rise to the sequence $x_0$, $x_1$, $x_2,...$. Under some mild conditions, the sequence converges to  the fixed point $\bar x$~\cite{cegielski2012iterative}. Hence, the values of the force and frequency that  maximize the selectivity are retrieved, as detailed in Algorithm~\ref{alg:algorithm1}. We note that the same procedure is applied to maximize selectivity in the case of protein population.
 
\begin{algorithm}
\caption{Joint Optimization using Fixed-Point Iteration}\label{alg:algorithm1}
\begin{algorithmic}[1]
\STATE Initialize the max number of iterations $I_{max}$ and the allowed error tolerance $\epsilon$. Set the iteration index $k=0$. Initialize the  force $f_{0}$ and the frequency $\omega_{0}$
\REPEAT
\STATE {$f_{k+1}=g_{f}(f_k)$ and $\omega_{k+1}=g_{\omega}(\omega_k)$}
\IF {$|f_{k+1}-f_k|<\epsilon$ and  $|\omega_{k+1}-\omega_k|<\epsilon$,}
\STATE $f^{*}=f_{k+1}$
\STATE $\omega^{*}=\omega_{k+1}$
\ELSE
\STATE  $f_k=f_{k+1}$
\STATE $\omega_k=\omega_{k+1}$       
\ENDIF
\UNTIL $k=I_{max}$
\STATE Output: $f^{*}$, $\omega^{*}$
\end{algorithmic}
\end{algorithm}

\section{Numerical Results}
\label{Sec:Sec5}

In this section,  we demonstrate the results of numerically simulating our analytical model. For our analysis, we will consider several cases of protein populations in the vicinity of one other to mimic different intra-body scenarios. Those proteins include rhodopsin, bacteriorhodopsin and D96N bacteriorhodopsin. The  aforementioned proteins  serve as excellent model systems for studying  properties of membrane proteins. Nonetheless,   although they contain seven $\alpha$-helical transmembrane domains, each has a different topology. The description of the normal mode motion and its associated vibrational frequency   for each of those proteins has been obtained from~\cite{balu2008terahertz}, while their mechanical properties  were found in~\cite{sapra2008mechanical} as summarized in Table~\ref{table:experimental}. It is to be noted that although the frequencies we will focus  on in the numerical results are those of rhodopsin ($1.36$~THz) and bacteriorhodopsin ($1.13$~THz), protein vibrational frequencies span the range between  $0.03$-$6.0$~THz\cite{mackerell1998all}. In addition, unless otherwise stated, we consider a protein population of $n=1000$. 
 
\begin{table*}[!]
\caption{ Protein Experimental Values}\begin{center}
\centering
\small
\label{table:experimental}
\begin{tabular}{|p{82pt}|p{36pt}|p{55pt}|p{48pt}|p{250pt}|}\hline
\begin{footnotesize}\textbf{Protein} \end{footnotesize}&\begin{footnotesize} \textbf{Frequency} (THz)\end{footnotesize}& \begin{footnotesize} \textbf{Stiffness} (N/m)  \end{footnotesize} & \begin{footnotesize} \textbf{Mass} (kg)  \end{footnotesize} &\begin{footnotesize}\textbf{ Conformational Change Associated with  Vibrational Mode} \end{footnotesize}  \\\hline
 \begin{footnotesize}Rhodopsin  \end{footnotesize}&\begin{footnotesize} 1.36~\cite{balu2008terahertz} \end{footnotesize} & \begin{footnotesize}3~\cite{sapra2008mechanical} \end{footnotesize} & \begin{footnotesize} $1.62\times10^{-24}$ \end{footnotesize}& \begin{footnotesize}Extracellular loops and helical region near extracellular surface move outward~\cite{balu2008terahertz}\end{footnotesize}.\\\hline
 \begin{footnotesize}Bacteriorhodopsin \end{footnotesize}& \begin{footnotesize}1.13 ~\cite{balu2008terahertz}   \end{footnotesize}& \begin{footnotesize}1.9~\cite{sapra2008mechanical}\end{footnotesize} &\begin{footnotesize} $1.48\times10^{-24}$ \end{footnotesize}& \begin{footnotesize}Extracellular loops movement: Helix A (in-out); Helix B (in-out); Helix C (up-down); Helix D (in-out); Helix F (out)~\cite{balu2008terahertz}.\end{footnotesize}\\\hline
\begin{footnotesize}D96N Bacteriorhodopsin \end{footnotesize}& \begin{footnotesize}1.025~\cite{balu2008terahertz}   \end{footnotesize}& \begin{footnotesize}Calculated using  $\omega_o^2=k/m$. \end{footnotesize} &\begin{footnotesize} $1.48\times10^{-24}$ \end{footnotesize}& \begin{footnotesize}Entire protein moves in-out
of membrane~\cite{balu2008terahertz}.\end{footnotesize}\\\hline
\end{tabular}
\end{center}
\end{table*}
\subsection{ Protein Response to Nanoantenna}
\label{Sec:part1}
\begin{figure}[htp]
\centering
\subfigure[]{%
  \includegraphics[height=4.2 cm, width=7cm]{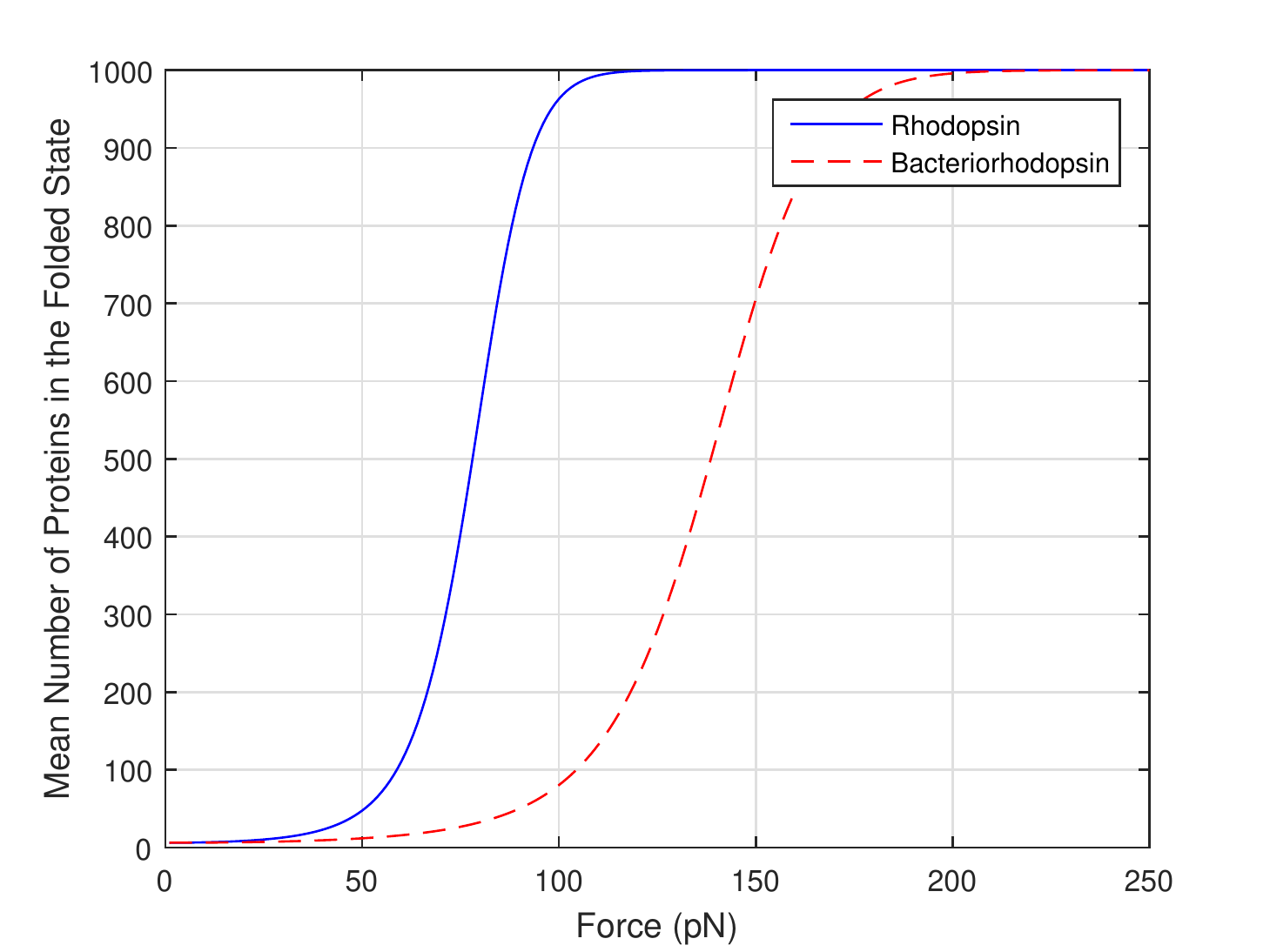}%
}
\subfigure[]{%
  \includegraphics[height=4.2 cm, width=7cm]{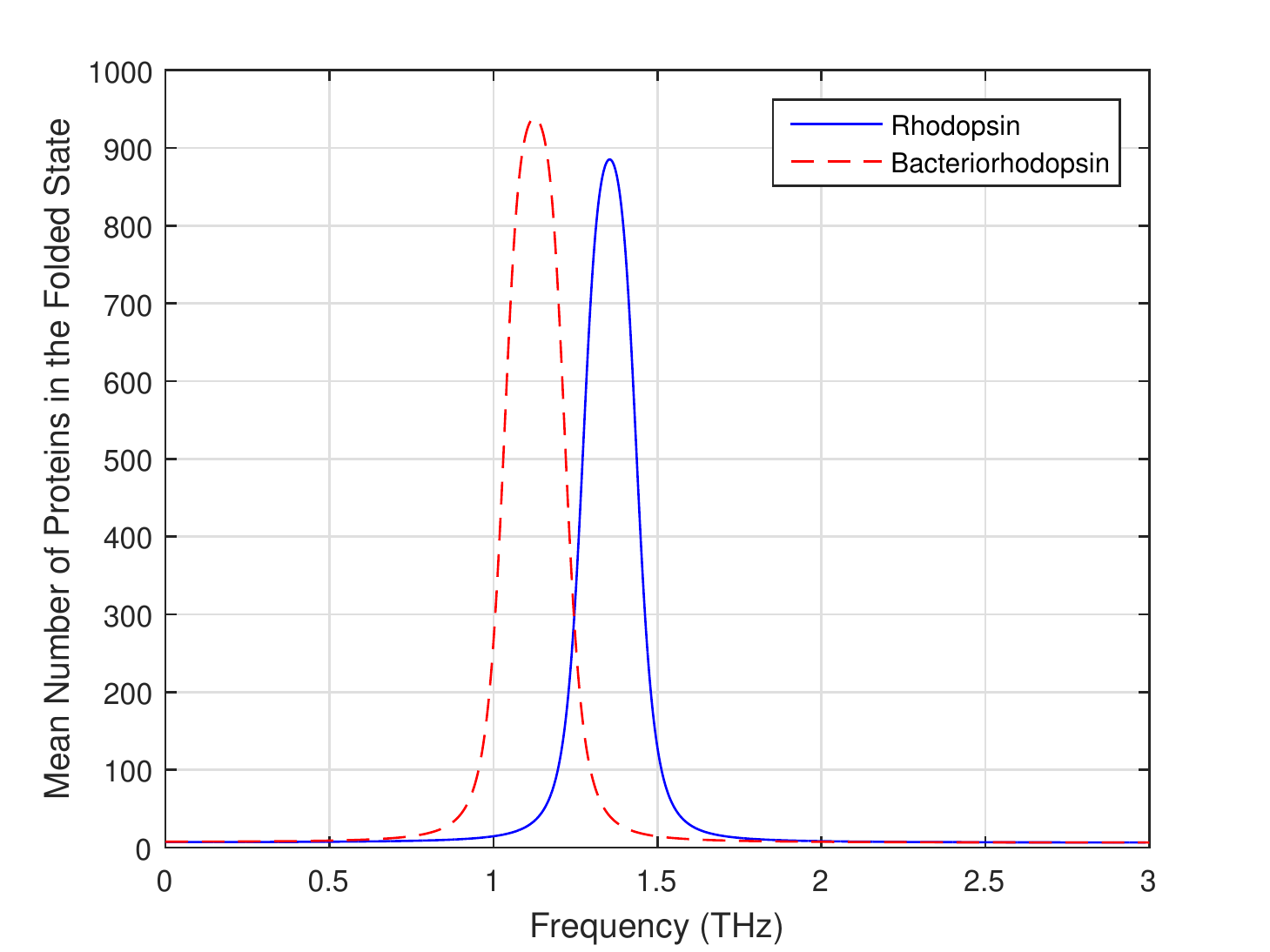}%
}

\caption{ (a) Mean number of folded proteins versus force. (b) Mean number of folded proteins versus frequency.}
\label{fig:1}
\end{figure}

Fig.~\ref{fig:1}a presents the expected number of proteins in the folded state, $E[n_F] =
np_F$, versus the nanoantenna  force  for both rhodopsin and bacteriorhodopsin, respectively.  The  driving frequency is fixed to the resonant frequency of rhodopsin (i.e. 1.36  THz), while the damping value  is  set to 0.3 THz satisfying $\omega_o>\frac{\gamma}{2}$. As indicated in the discussion,  the presence of such underdamped delocalized modes in proteins  has significant implications for the understanding of the efficiency of ligand binding, protein-molecule interactions, as well as biological functions~\cite{turton2014terahertz}.
Moreover, the free energy value, $\Delta E$, is set to 6 $k_bT$ since this value is considered as a lower bound for the free energy difference between two states~\cite{benham2009mathematics}.

  The folded state of the protein becomes populated when tuning the nanoantenna frequency to the natural vibrational frequency  of the desired protein. This shows that the system gets driven when subjected to the nanoantenna  force. For the other untargeted protein,  there is a probability that it acquires a folded state  if an excessive amount of force is applied to the system. This is why the nanoantenna force should be chosen within a specific window and should not exceed a particular threshold to ensure that it only impacts  the protein population of interest. This aspect will be discussed in depth in Sec.~\ref{Sec:joint}.

Fig.~\ref{fig:1}b illustrates the expected  number of folded proteins versus frequency for both rhodopsin and bacteriorhodopsin.  For this case, the value of the nanoantenna force is fixed to $100~$pN, while the damping value is set to $0.3$~THz.  The presented results reveal that by tuning the nanoantenna frequency to the resonant frequency of the protein, the probability of being in a folded state is maximum. As such, the protein can acquire the state which induces the desired biological function. We also observe that the probability of the folded state of bacteriorhodopsin is higher in comparison to rhodopsin since as seen in Table~\ref{table:experimental} it has both  lower stiffness and mass. This leads to a higher value of the average driving energy, $\left\langle E_{\mathrm{tot}}\right\rangle_{ss}$, indicating that  the   protein mechanical  system plays a role  in controlling the protein kinetic rates.

\subsection{Selectivity  of the Nanoantenna-Protein Interaction}

\subsubsection{Resonant Interactions}

\begin{figure}[h!]
\centering
\includegraphics[height=4.2 cm, width=7cm]{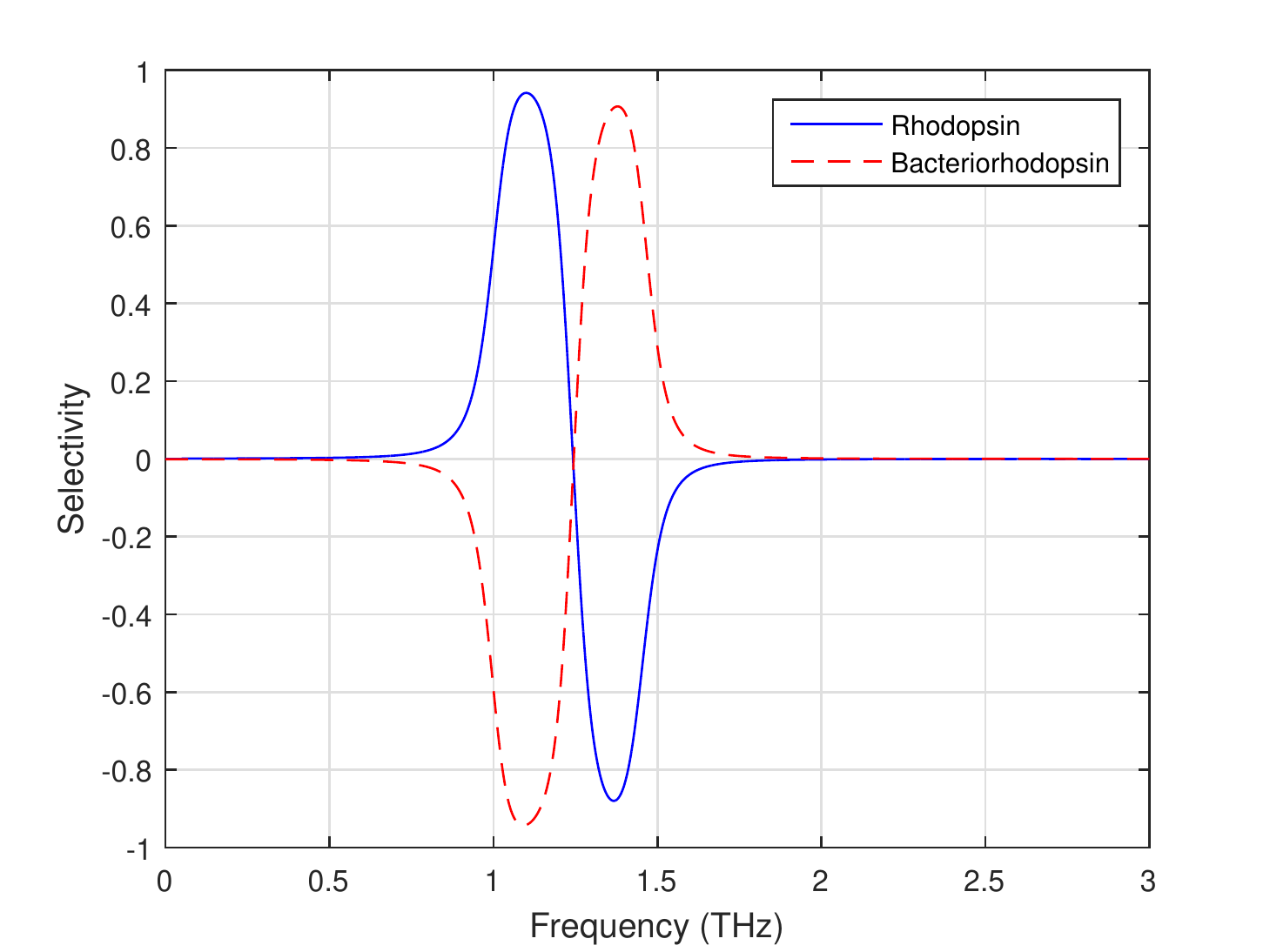}
\footnotesize
\caption{Selectivity values for the rhodopsin and bacteriorhoodopsin
population versus frequency.} 
 \label{fig:combined}
 \end{figure}

To evaluate the interaction between the nanoantenna and the targeted protein population, we need to compute the selectivity of the system under different scenarios. For all our simulations, we  use a force value  of $f_o=100~$pN and a damping value of $\gamma=0.3$~THz. Fig.~\ref{fig:combined} demonstrates  the   selectivity  versus frequency when the targeted protein population is either rhodopsin or bacteriorhodopsin, respectively. We see that, as expected, with equal populations,  the interaction achieves  maximum selectivity when the nanoantenna is set to the resonant frequency of the desired protein population. 

The selectivity value of rhodopsin is $0.9$ at $1.36$~THz, while that of bacteriorhodopsin is $0.93$ at $1.13$~THz. The slight difference between the selectivity values of the aforementioned  protein populations stems from the fact that the folding probability of bacteriorhodopsin is higher than that of rhodopsin due to its mechanical structure (as explained in Sec.~\ref{Sec:part1}). The presented selectivity results indicate that a high number of proteins of the targeted  population has been folded at resonance, with very low false positives.   Hence, the THz frequency  provides  resonant  access  to  protein fundamental  modes allowing experimenters  to probe molecular responses with high selectivity. The possible outcomes of a nanoantenna-protein interaction are summarized in Table~\ref{table:confusion_matrix}.
\begin{table*}[!]
\caption{ Possible Outcomes of a Nanoantenna-Protein Interaction}\begin{center}
\centering
\small
\label{table:confusion_matrix}
\begin{tabular}{|p{95pt}|p{90pt}|p{90pt}|}\hline
\begin{footnotesize} \end{footnotesize}&\begin{footnotesize} Folded Protein State\end{footnotesize}& \begin{footnotesize} Unfolded Protein State \end{footnotesize}  \\\hline
 \begin{footnotesize}Targeted Population  \end{footnotesize}&\begin{footnotesize}True Positive \textbf{(TP)} \end{footnotesize} & \begin{footnotesize}False Negative (\textbf{FN})\end{footnotesize}\\\hline
 \begin{footnotesize}Untargeted Population\end{footnotesize}& \begin{footnotesize}False Positive \textbf{(FP)}  \end{footnotesize}& \begin{footnotesize}True Negative(\textbf{TN})\end{footnotesize} \\\hline
\end{tabular}
\end{center}
\end{table*}

  Moreover, at the resonant frequency of the untargeted protein population,  a negative selectivity value is attained. As illustrated in Sec.~\ref{Sec:Sec4}, a  negative  value signifies that the untargeted protein population existing in the system is selected. This  acts as a pre-indication for experimenters to avoid getting close to this frequency as it will lead the undesired  population to acquire a folded conformation.
  
  Finally, the region of zero crossing  in Fig.~\ref{fig:combined} designates no selectivity. At these frequencies, the mean number of folded proteins of the existing populations is very close to one another, and hence neither population  experiences a conformational change.

\begin{figure}[h!]
\centering
\includegraphics[height=4.2 cm, width=7cm]{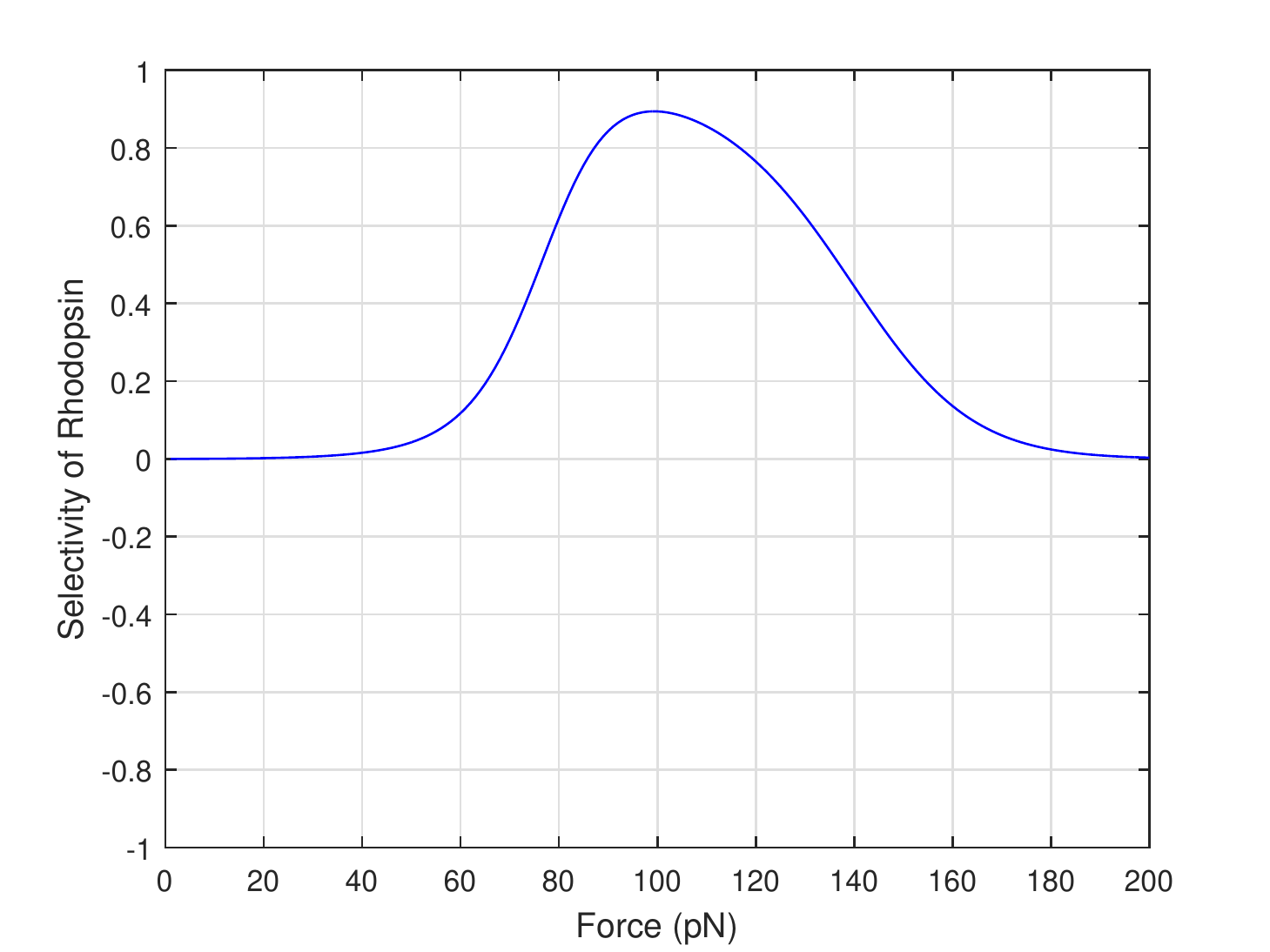}
\footnotesize
\caption{Selectivity values when targeting the rhodopsin
population versus force.} 
 \label{fig:4}
 \end{figure}

An important  question that arises within this context is how to further achieve a higher selectivity
value at resonance. Recall, in our system, a number of parameters act as controlling factors. Among others, the nanoantenna force plays a fundamental role as its value is determined by the experimenter. Fig.~\ref{fig:4} illustrates the selectivity of the rhodopsin population  versus the nanoantenna force at its resonant frequency. We see that  the nanoantenna force value must be chosen within a specific range to deem the interaction selective. A very low or very high force value will result in zero selectivity since the force will either be too low to fold the protein, or too high making all the proteins in the system fold.

The selected force value depends on several factors including the protein type and system characteristics. For instance, the higher the damping, the more viscous the system is and as such more force is required. Damping provides insights into the dynamics of protein conformational changes by demonstrating the relative significance of frictional forces in activating reactions. Although we cannot control damping as it is dictated by the system, it is beneficial to study its effect on selectivity. 

\begin{figure}[h!]
\centering
\includegraphics[height=4.2 cm, width=7cm]{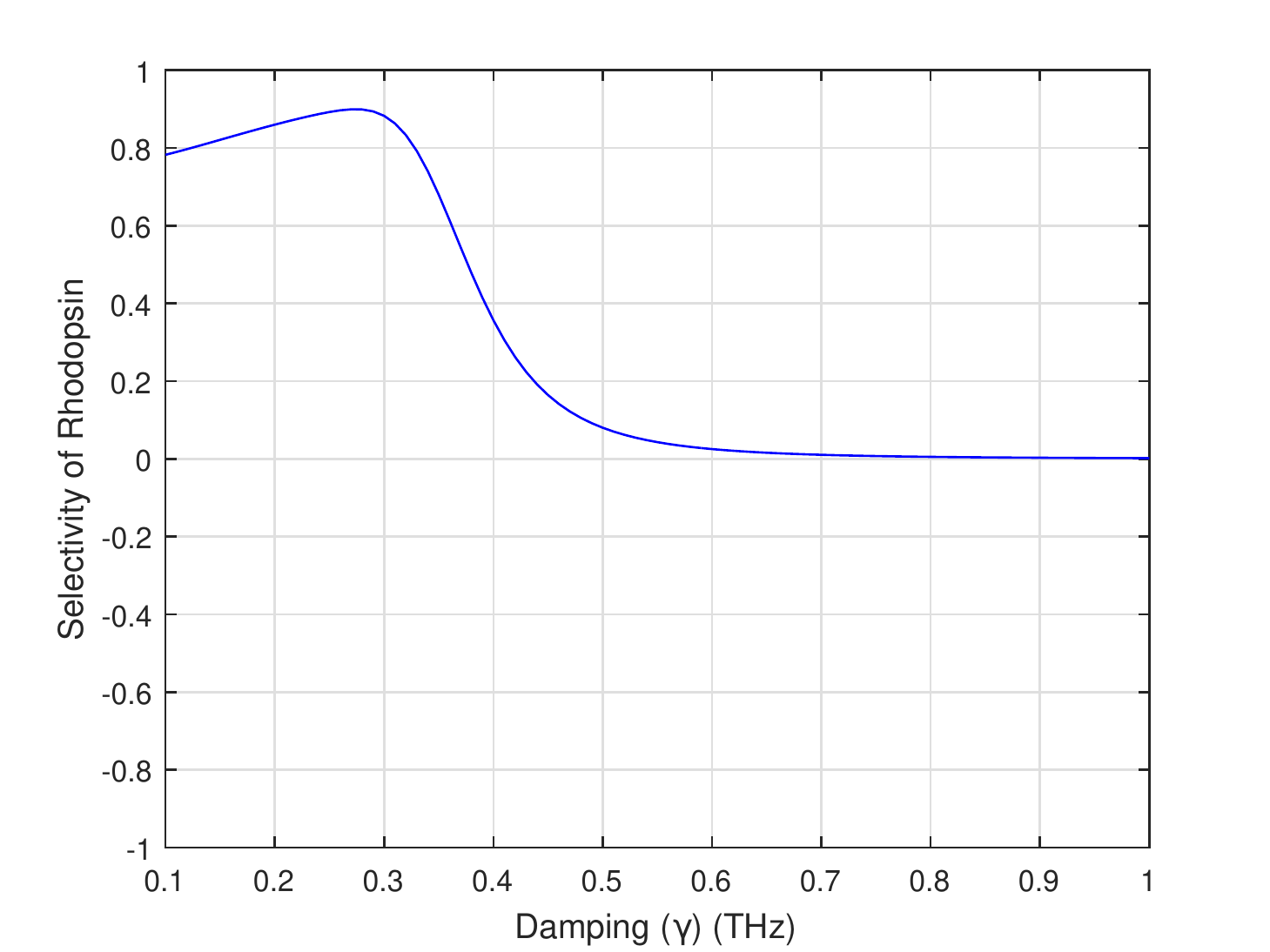}
\footnotesize
\caption{Selectivity values when targeting the rhodopsin
population versus damping, $\gamma$.} 
 \label{fig:selectivity_gamma}
 \end{figure}

Fig.~\ref{fig:selectivity_gamma} demonstrates the selectivity of the rhodopsin population  versus damping. The force is set to $100$~pN and we are still operating at resonance. Similar to the force, damping  must be within a certain range for the interaction to be considered selective. The system cannot sustain a high damping value since in this case the viscous forces will  impact the capability of identifying the THz modes associated with conformational changes. In other words, the distinct vibrational modes will be faded by the frictional forces and the resonant effect will die out. Nonetheless, this should not be an issue since structured bio-macromolecules such as proteins in  the THz frequency  range are underdamped, satisfying the condition in which $\omega_o>\frac{\gamma}{2}$~\cite{turton2014terahertz}. 

\subsubsection{Multiple Proteins}
\begin{figure}[htp]
 \centering
\subfigure[]{%
  \includegraphics[height=4.2 cm, width=7cm]{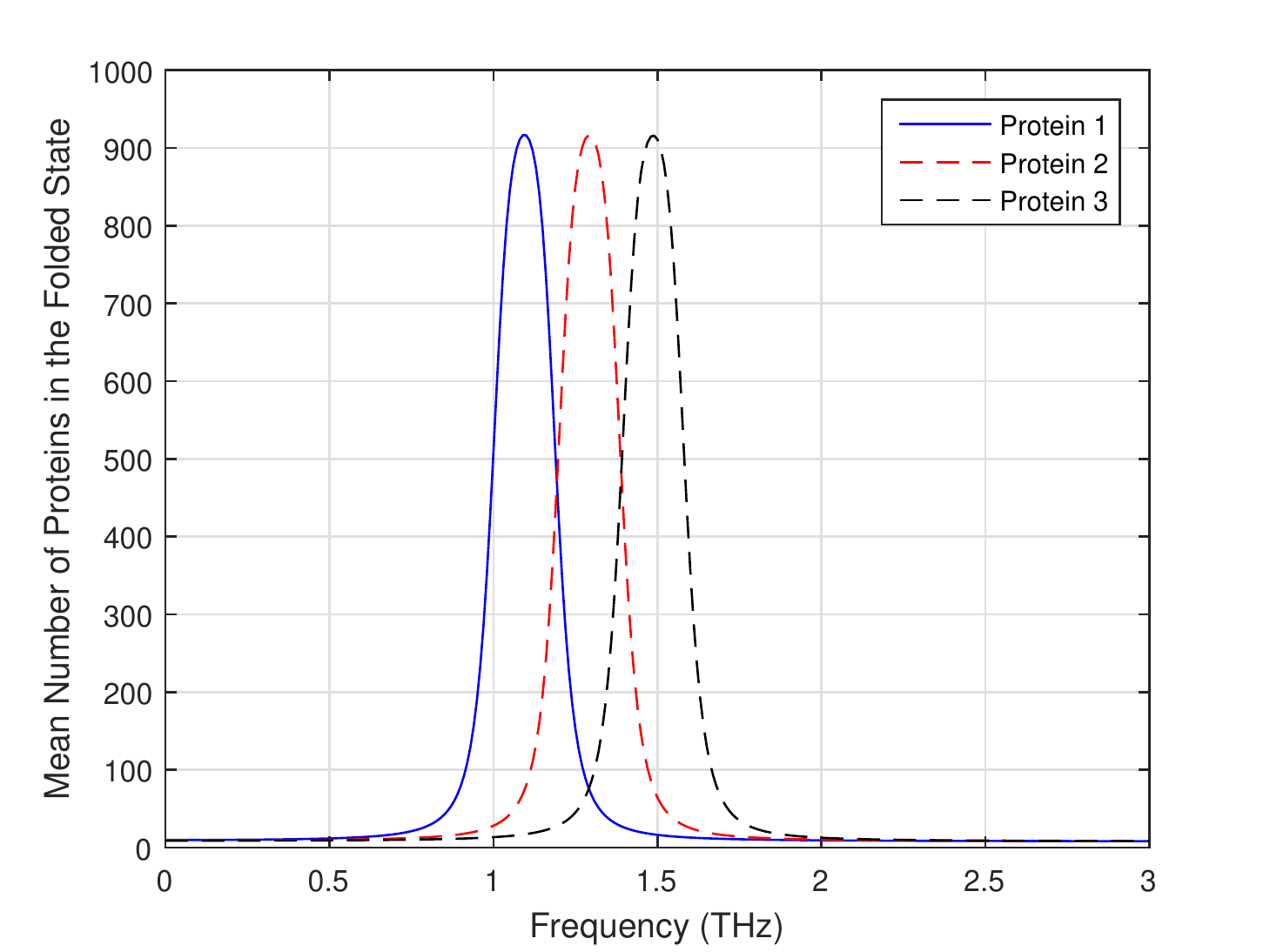}%
}
\subfigure[]{%
  \includegraphics[height=4.2 cm, width=7cm]{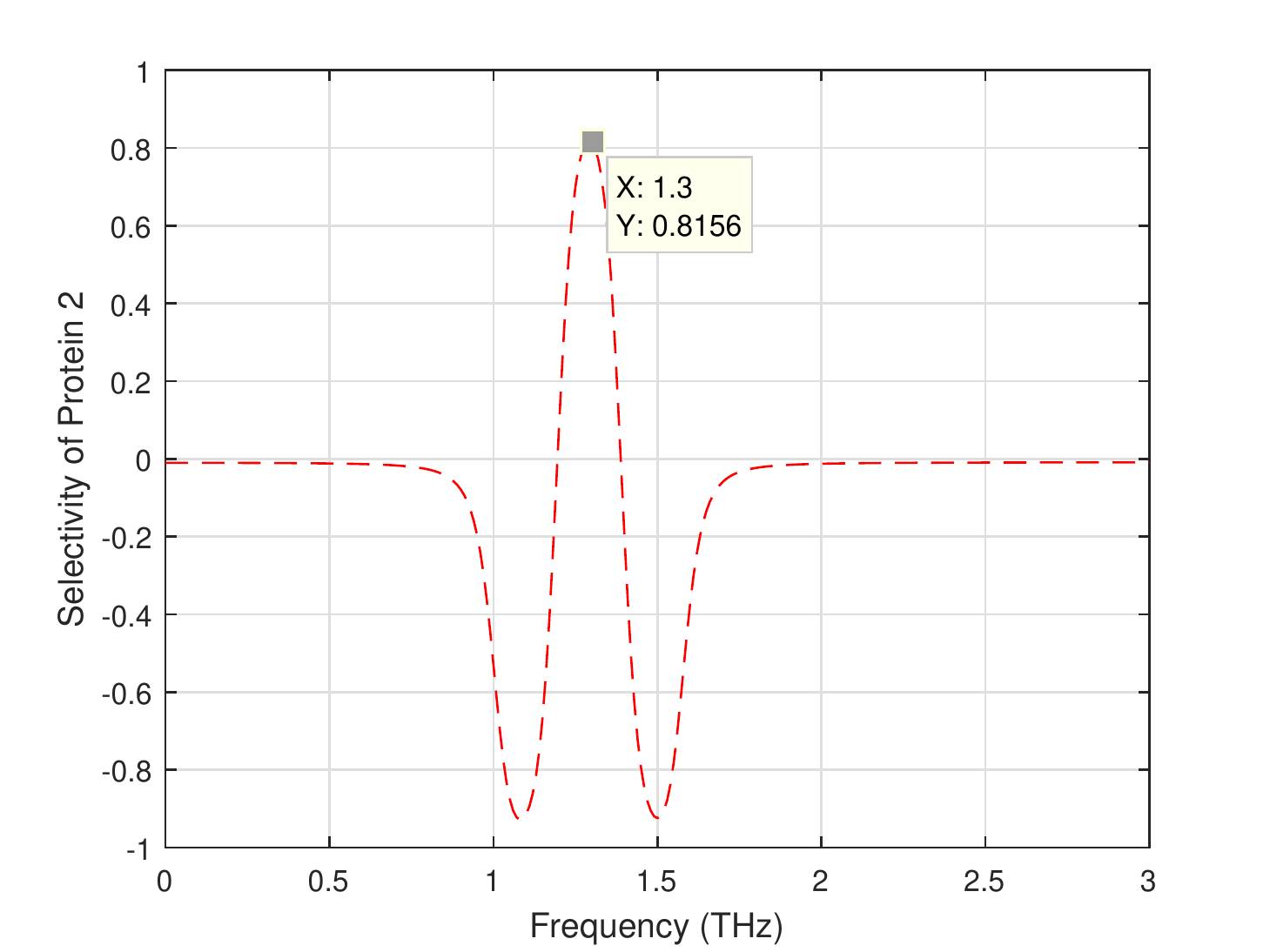}%
}

\caption{(a) Mean number of folded proteins versus frequency for protein population  1, 2 and 3.  (b) Selectivity values for protein 2 population versus frequency.}
\label{fig:3proteins}
\end{figure}

One important feature of the developed selectivity metric is its applicability to a  scenario involving multiple undesired proteins. Such an example fits applications where we are interested in targeting   specific disease associated cells  amongst  others. To illustrate this case, we choose three protein populations  with resonant frequencies of $1.1$, $1.3$ and $1.5$~THz, as demonstrated  in Fig.~\ref{fig:3proteins}a. The  protein population are of equal abundance, $n_1=n_2=n_3=1000$. We are interested in targeting protein population 2, with a resonant frequency of 1.3 THz. 

Fig.~\ref{fig:3proteins}b illustrates selectivity versus frequency.  It can be seen that the selectivity is maximum at $1.3$~THz, where the marker  shows that the selectivity value of protein 2 at resonance is $0.81$. Such a high value is an indicator that the  desired proteins can be discriminated from the other populations and therefore can be targeted by the nanoantenna. It also confirms the effectiveness of the proposed selectivity metric in capturing the performance of the system.

\subsubsection{Joint Optimization}
\label{Sec:joint}
\begin{figure}[h!]
\centering
\includegraphics[width=0.4\textwidth]{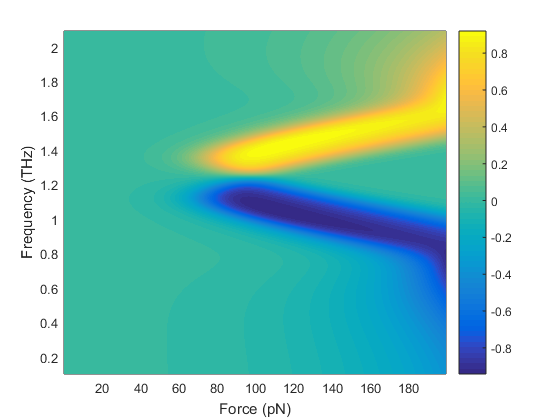}
\footnotesize
\caption{Pseudo-color plot showing the optimal force and frequency for targeting rhodopsin in the presence of bacteriorhodopsin.}  
\label{fig:pseudo1}
\end{figure}

\begin{figure}[h!]
\centering
\includegraphics[width=0.4\textwidth]{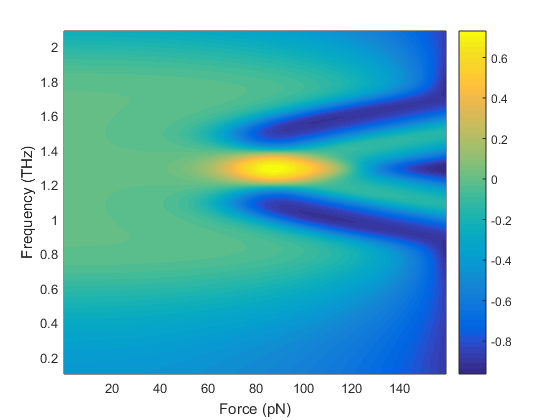}
\footnotesize
\caption{Pseudo-color plot showing the optimal force and frequency for targeting protein population 2 ($\omega=1.3$ THz) in the presence of population 1 ($\omega=1.1$ THz) and population 3 ($\omega=1.5$ THz).}
\label{fig:pseudo2}
\end{figure}

To design a system that achieves maximum selectivity, we  need to know the optimal values of both the nanoantenna force and frequency. Hence, we cast a joint optimization formulation for selectivity with respect to those parameters as explained in Sec.~\ref{Sec:Sec4a}. Fig.~\ref{fig:pseudo1} emulates the same scenario provided in  Fig.~\ref{fig:combined}. It presents a pseudo-color plot indicating the optimal force and frequency values that must be used when targeting protein rhodopsin in the presence of bacteriorhodopsin. The color map demonstrates the intensity of the selectivity value. Similarly, Fig.~\ref{fig:pseudo2} presents a pseudo-color plot, which mimics the scenario presented in Fig.~\ref{fig:3proteins}. It also provides the force and frequency values that result in obtaining a maximum selectivity when targeting protein population 2 in the presence of populations 1 and 3. 

We can notice from Figs.~\ref{fig:pseudo1} and~\ref{fig:pseudo2} that the joint optimization is a very powerful tool that provides experimenters with access to the optimal force and frequency values. As such, they can  guarantee that the system achieves maximum selectivity and that the targeted protein population acquires the desired folding behavior. The color map provides a visualization of   the number of protein populations considered in each system and  it clearly discriminates regions that should be targeted from those that should be avoided in order not to activate the wrong proteins. 
 
 \subsubsection{Shift in Optimal Frequency}

In Fig.~\ref{fig:combined}, we noticed that a maximum selectivity is always achieved at resonance. This occurs when the resonant frequency difference between the proteins involved is within a certain range. In fact, if the difference between the protein resonant frequencies is $\geq0.2$ THz, then selectivity is always maximum at resonance. This value has been attained after   conducting several simulation trials and experimenting with different proteins of varying resonances. Nevertheless, if the difference is less than $0.2$ THz, the overlap between the protein populations  gets   larger and more of the untargeted  protein population gets folded resulting in a higher number of false positives. Such an impact not only  lowers the  selectivity value but also affects the frequency at which the maximum response is achieved.

\begin{figure}[htp]
 \centering
\subfigure[]{%
  \includegraphics[height=4.2 cm, width=7cm]{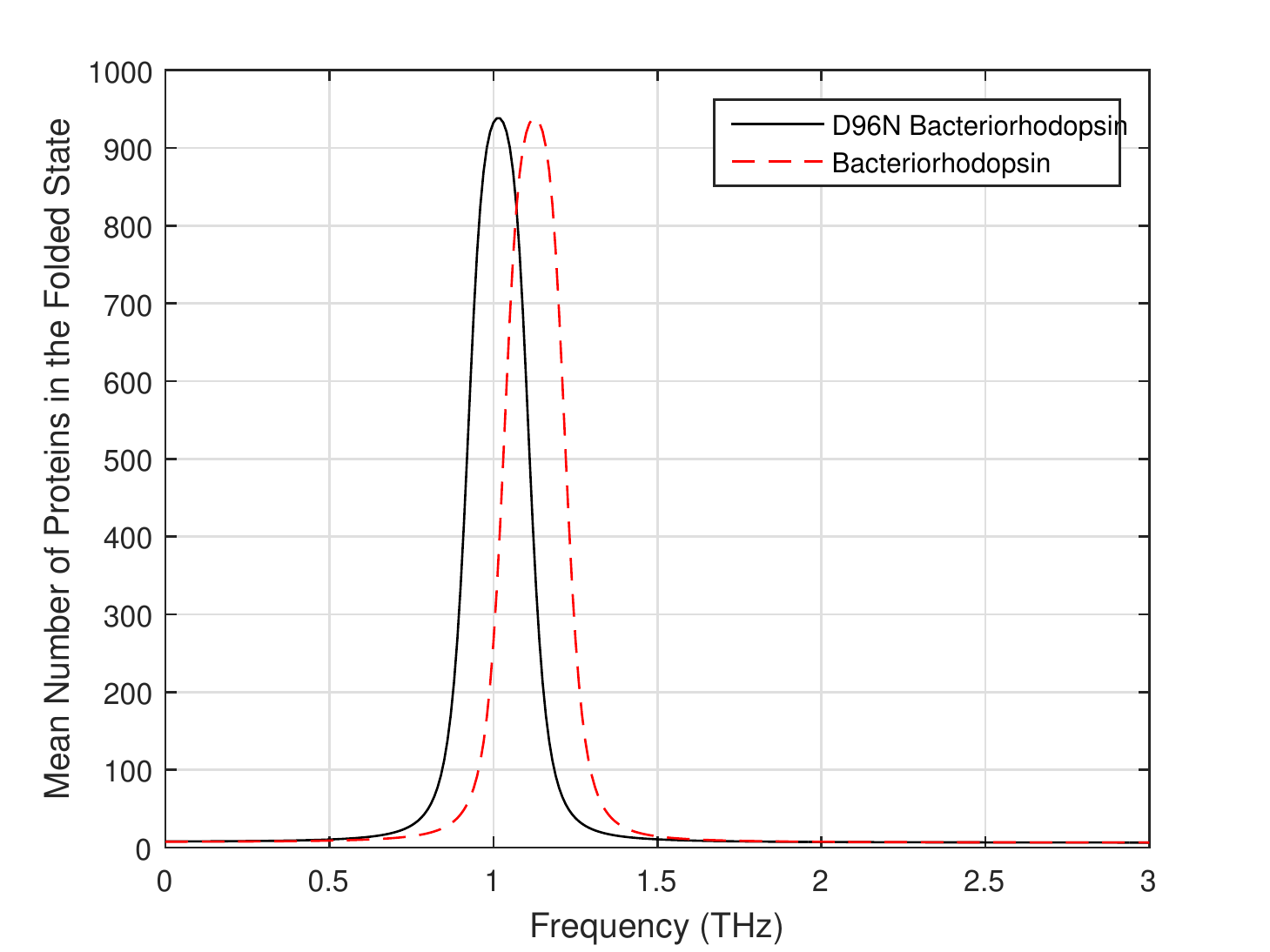}%
}
\subfigure[]{%
  \includegraphics[height=4.2 cm, width=7cm]{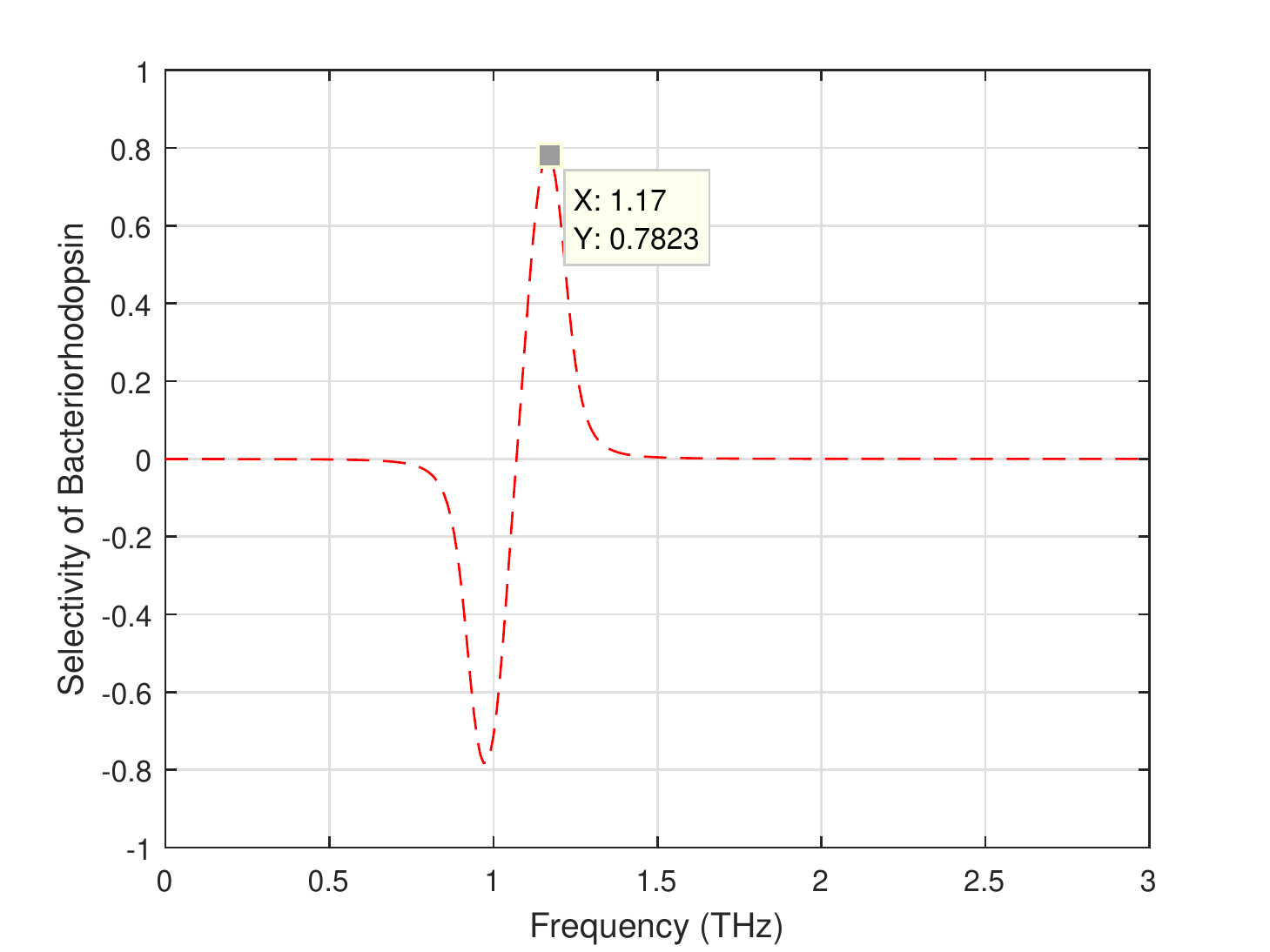}%
}

\caption{(a) Mean number of folded proteins versus frequency for bacteriorhodopsin and D96N bacteriorhodopsin. (b) Selectivity values for bacteriorhodopsin population versus frequency.}
\label{fig:selectivity_bacterio_D96}
\end{figure}

 Fig.~\ref{fig:selectivity_bacterio_D96}a presents the mean number of folded proteins for two protein populations with close resonant frequencies and equal abundance, namely, bacteriorhodopsin ($\omega=1.3$~THz) and D96N bacteriorhodopsin. The latter is a bacteriorhodopsin mutant with a resonant frequency of $1.025$~THz~\cite{balu2008terahertz}. The difference in their resonant frequency is~$\approx 0.1$~THz. This  results in a higher overlap between the protein populations. 

Fig.~\ref{fig:selectivity_bacterio_D96}b shows the selectivity of bacteriorhodopsin versus frequency in the presence of the D96N mutant. It can be seen that not only has the selectivity value decreased to $\approx 0.78$ in comparison to the selectivity of bacteriorhodopsin in Fig.~\ref{fig:combined},  but the maximum selectivity is achieved at $1.17$~THz rather than the expected resonant frequency of $1.13$~THz. Thereby, the selectivity metric is important since it indicates the frequency at which  the maximum system desired response  is achieved; this could be at the resonant frequency, $\omega_o$, or at a frequency close to it.  

If we considered an imbalance in population, in which for example $n_1=800$ proteins, while $n_2=1000$ proteins, the   selectivity value is going to  decrease. This occurs  since the number of proteins of the targeted population has decreased, resulting in a lower  number of folded proteins.  The countereffect occurs if the number of proteins in the targeted population is larger than the  untargeted  one, where the selectivity value is going to increase.

\section{Conclusions} 
\label{Sec:Sec6}
In this paper, we derive an expression for the selectivity of the nanoantenna-protein response. Selectivity serves as a metric  that allows us to study whether the interaction  between the nanoantenna and the protein is sufficient to  provoke a conformational change in the desired protein molecule/population in the presence of other proteins. The metric provides a score that ranges between $-1$ and $1$. A value of $-1$ indicates the worst-case scenario, where only the  undesired protein  molecule/population is  being provoked to fold. On the contrary, a value of $1$ indicates the best-case scenario, where only the desired protein molecule/population is being stimulated to fold. As such, the closer the  value achieved to is $1$, the higher the selectivity in the system.  According to the  selectivity values attained, an experimenter can determine the level of control over the desired nanoantenna-protein  interaction.
The effectiveness of the proposed metric is seen from the different tested scenarios.  

To develop the selectivity metric, we deploy the Langevin stochastic equation driven by an external force  to capture the protein dynamics in an intra-body environment.   We  then formulate an expression for the steady-state energy stimulating the protein to change its conformation and use  it to amend Boltzmann distribution. For the numerical analysis, we consider well-studied proteins with experimental parameters available in the literature. The achieved results provide an understanding of  the impact of the system parameters, namely, the nanoantenna force, damping, and protein abundance on the performance. Such  analysis provides genuine  insight into the conditions at which maximum selectivity is achieved.

The presented work provides a novel perspective regarding signal transduction processes, where proteins are controlled via EM waves to obtain their functional conformation as part of a vibratory network. By understanding  how THz waves may alter the pathological mechanisms in intra-body networks, the development of  therapeutic strategies shall be enabled and wide-scale  shift by pharmaceutical research is motivated.   
\bibliographystyle{IEEEtran}

\end{document}